\documentclass[twocolumn,floatfix]{revtex4-2}

\usepackage{graphicx}
\usepackage{dcolumn}
\usepackage{bm}
\usepackage{tabularray}
\usepackage{array}

\usepackage{amsmath}
\usepackage{xcolor}

\usepackage{hyperref}
\usepackage{cleveref}

\usepackage{subfig}

\definecolor{nblue}{RGB}{28,130,185}
\definecolor{cgreen}{RGB}{76,153,0}
\definecolor{myorange}{RGB}{245,156,74}

\graphicspath{{Illustrations}{Results}}

\hypersetup{
  colorlinks=true,
  citecolor=cyan,
  urlcolor=cgreen
}

\begin{document}
\preprint{APS/123-QED}
\title{Obstacle-aware navigation of smart microswimmers in a turbulent flow }

\author{Vaishnavi Gajendragad}
\email{vaishnavi.gajendragad@ds.mpg.de}
\affiliation{Max Planck Institute for Dynamics and Self-Organization, G$\ddot{o}$ttingen, Germany.}
\author{Akanksha Gupta}%
\email{guptaakanksha17@gmail.com}
\affiliation{Department of Physics, Maulana Azad National Institute of Technology (MANIT), Bhopal - 462003, India}
 
 \author{Nadia Bihari Padhan}
\email{nadia\_bihari.padhan@tu-dresden.de}
\affiliation{Institute of Scientific Computing, TU Dresden, 01069 Dresden, Germany}
\author{Rahul Pandit }
\email{rahul@iisc.ac.in }
\affiliation{Centre for Condensed Matter Thoery, Department of Physics, Indian Institute of Science, Bangalore - 560012, India }

\begin{abstract}
\noindent
Microswimmers in turbulent flows often navigate complex, heterogeneous, and obstacle-rich environments, where they exhibit intricate behaviors such as trapping at and escape from obstacles. 
We generalize recent $\mathcal{Q}-$learning methods of J.K. Alageshan \textit{et al.} [Phys.Rev.E \textbf{101}, 043110 (2020)] and A. Gupta \textit{et al.} [Physics of Fluids \textbf{37}, 045107 (2025)] developed for non-interacting microswimmers that aim to move optimally from an initial position to a target, to account for the additional complication of an obstacle in the flow. We begin by considering one circular obstacle in forced two-dimensional (2D) Navier-Stokes turbulence in which the energy spectrum displays a forward cascade. We employ the volume-penalization method to introduce this obstacle within our doubly periodic simulation domain. We augment our adversarial $\mathcal{Q}-$learning \cite{Alageshan_2020, gupta2024flockingaidpathplanning} by suppressing the tendency of microswimmers to get trapped in stagnation points in the vicinity of the obstacle. We demonstrate that smart microswimmers ($SS$), which adopt our obstacle-aware adversarial $\mathcal{Q}-$learning strategy, outperform both na\"ive swimmers ($NS$) and surfers ($SuS$). 
\end{abstract}

\maketitle  

\section{\label{sec:level1}Introduction}
\noindent
Motility has played a crucial role in the evolutionary success of microorganisms. Not only has it enabled them to navigate effectively through complex environments, to reach sources of nutrition, but it has also helped them to avoid obstacles.  Sperm transport provides an interesting example: millions of spermatozoa face a series of obstacles in fluctuating fluid environments, but using head-based chemotaxis, sperm cells adjust their propulsion in response to environmental chemical gradients to enhance their chances of successful fertilization~\cite{10.1093/humupd/dmi047}. Similarly, bacteria like \textit{Escherichia coli} employ a \textit{run and tumble} strategy to explore their surroundings, in the digestive tract, where they assess their proximity to attractants like sugars and amino acids. When a repellent is detected, \textit{E. coli} increase the frequency of tumbles, changing direction so as to move away from harmful stimuli~\cite{Berg_2005, Tailleur_2008, alegado_king_2014_bacterial}.
\newline
\noindent
Swimmer navigation in steady flows has been well documented~\cite{Schneider_Stark_2019, Michalec_2015, Elgeti_2015, kiverin2025twophasehydrodynamicmodelactive} and studied in biophysics. Equally important is the path planning of artificial microswimmers or microrobots
through fluid flows; if these flows are turbulent, then this path-planning problem is challenging, but some progress has been made recently using reinforcement learning~\cite{Alageshan_2020,biferale2019zermelo,gupta2024flockingaidpathplanning}. The generalization of such reinforcement-learning-aided path planning in obstacle-laden environments is an open problem. We address this here.

Experiments on microscopic self-propelled particles, in patterned environments, have shown that they can navigate complex spatial features, particularly near walls, where they first make contact, slide along the surface while rotating, and eventually reorient their propulsion direction to detach and move away~\cite{Volpe_2011}. Additionally, Image-Based Visual Servoing methods have been successfully employed to train helical microswimmers to follow a designated path and effectively avoid obstacles~\cite{Xu_2020}. Studies on the role of hydrodynamic interactions of active microswimmers with walls and obstacles have observed that optimal path planning is counterintuitive because the flow field generated by their motion \cite{Daddi-Moussa-Ider_2021} can result in bound states, because of circling or trapping~\cite{Lauga2006, Ramprasad_PRE2025}; furthermore, they can be captured or scattered depending on  hydrodynamic interactions~\cite{Spagnolie2015}.

 At microscopic scales, inertia is negligible, so swimmer dynamics is strongly influenced by viscosity and thermal fluctuations. However, navigation strategies that are effective in the viscous regime often fail in flow environments with strong shear, vortical structures, and turbulence. Over the past decade, machine-learning strategies have been developed for the path-planning of microswimmers in turbulent flows~\cite{Colabrese_2017,Gustavsson_2017,Colabrese_Gustavsson_Celani_Biferale_2018,Alageshan_2020,Chakraborty2025SmartNavigation,Qiu2022Symmetry,gupta2024flockingaidpathplanning, ZOU2024107666}; these studies use different versions of reinforcement learning. Recent studies have shown that smart microswimmers can learn to exploit flow gradients and shear to migrate efficiently across Poiseuille flows, significantly outperforming na\"ive or passive strategies \cite{Chakraborty2025SmartNavigation}.  Reinforcement learning can be used to tame Lagrangian chaos and suppress dispersion in time-dependent flows by learning multi-objective control strategies \cite{Calascibetta2022Taming}. Related studies demonstrated that microswimmers can successfully navigate complex vortical and turbulent-like flows by exploiting transient flow features, rather than attempting to overcome them. Furthermore, reinforcement learning has been applied to undulatory microswimmers to demonstrate that gait modulation can be optimized to achieve efficient steering and transport in flow fields~\cite{Khiyati2023Steering}. Thus, intelligent navigation has emerged as a rapidly developing research frontier for microswimmers, both artificial and biological~\cite{Mo2023IntelligentMicroswimmers, Xue2024NanoMotors}; and there has been a concomitant growth in the application of machine learning (ML) to this field. Similar ideas have also been explored at larger spatial scales, e.g., in the navigation of robotic fish in vortical flows~\cite{Feng2024RoboticFish}. 
Moreover, it has been demonstrated~\cite{Qiu2022Symmetry} that successful navigation in steady flows depends crucially on symmetry breaking: when only local hydrodynamic cues are available, swimmers can learn efficient migration strategies provided that either the dynamics or the task itself breaks the underlying symmetries of the flow. 

Despite these significant advances, an important aspect of realistic microswimmer navigation remains largely unexplored: the effects on path planning  of solid obstacles in turbulent or complex flows. In many practical settings -- such as porous media, vascular networks, or cluttered microfluidic devices -- microswimmers~\cite{Li2009,Spagnolie2012,Chepizhko2013,AlonsoMatilla2019} must cope not only with flow-induced fluctuations, but also detect, avoid, and exploit obstacles. Obstacles fundamentally alter local-flow topology, generating boundary layers, recirculation zones, and intermittent high-shear regions that affect swimmer trajectories strongly. In turbulent flows, the interplay between unsteady vortical structures and obstacle-induced flow heterogeneity amplifies the complexity of navigation. Therefore, the development of ML strategies for microswimmer path planning, in turbulent flows with obstacles, is an important challenge; furthermore, it has applications in targeted drug delivery and the
locomotion of microbots~\cite{manish2011targeted, manzari2021targeted, singh2009nanoparticle}. Therefore, we investigate how microswimmers can learn adaptive strategies that account simultaneously for turbulent fluctuations and obstacle-induced flow features. 

In particular, we generalize  the $\mathcal{Q}-$learning methods of \cite{Alageshan_2020,gupta2024flockingaidpathplanning} for non-interacting microswimmers as follows: we employ slave microswimmers to implement an adversarial $\mathcal{Q}-$learning to optimize their path planning. Such microswimmers aim to move optimally, from an initial position to a target, \textit{while swimming around obstacles in the flow}. We begin by considering one circular obstacle in a forced 2D Navier-Stokes fluid; the forcing is chosen to induce turbulence in which the energy spectrum displays a forward cascade~\cite{boffetta2012two,Pandit_2017}; we use the volume-penalization method [see, e.g., \cite{schneider2005numerical}] to introduce this obstacle within our doubly periodic simulation domain. We augment our adversarial $\mathcal{Q}-$learning by suppressing the tendency of microswimmers to get trapped in stagnation points in the vicinity of the obstacle. Finally, we demonstrate that smart swimmers that adopt our obstacle-aware adversarial $\mathcal{Q}-$learning strategy outperform both na\"ive swimmers ($NS$) and surfers ($SuS$); the latter optimize path planning by using velocity-gradient information. Our results, which are the first of their kind for turbulent flows with an obstacle, show that, over time, the cumulative number $N^{SS}$ of smart microswimmers reaching the target increases significantly, after adversarial $\mathcal{Q}-$learning, relative to their na\"ive or surfer counterparts.
\noindent 
Some groups have examined the microswimmers and their path planning
via actor-critic reinforcement learning in two-dimensional (2D) turbulent flows~\cite{mecanna2025critical, koh2025physics,biferale2019zermelo}. For the gravitaxis of microswimmers, it has been shown
that surfers~\cite{monthiller2022surfing},  which follow velocity gradients in the flow, can outperform microswimmers that use the path-planning strategy from \cite{Alageshan_2020, gupta2024flockingaidpathplanning}. To the best of our knowledge, such studies have not been carried out with obstacles.
\begin{figure}[h!]
    \centering
    \includegraphics[width=\linewidth]{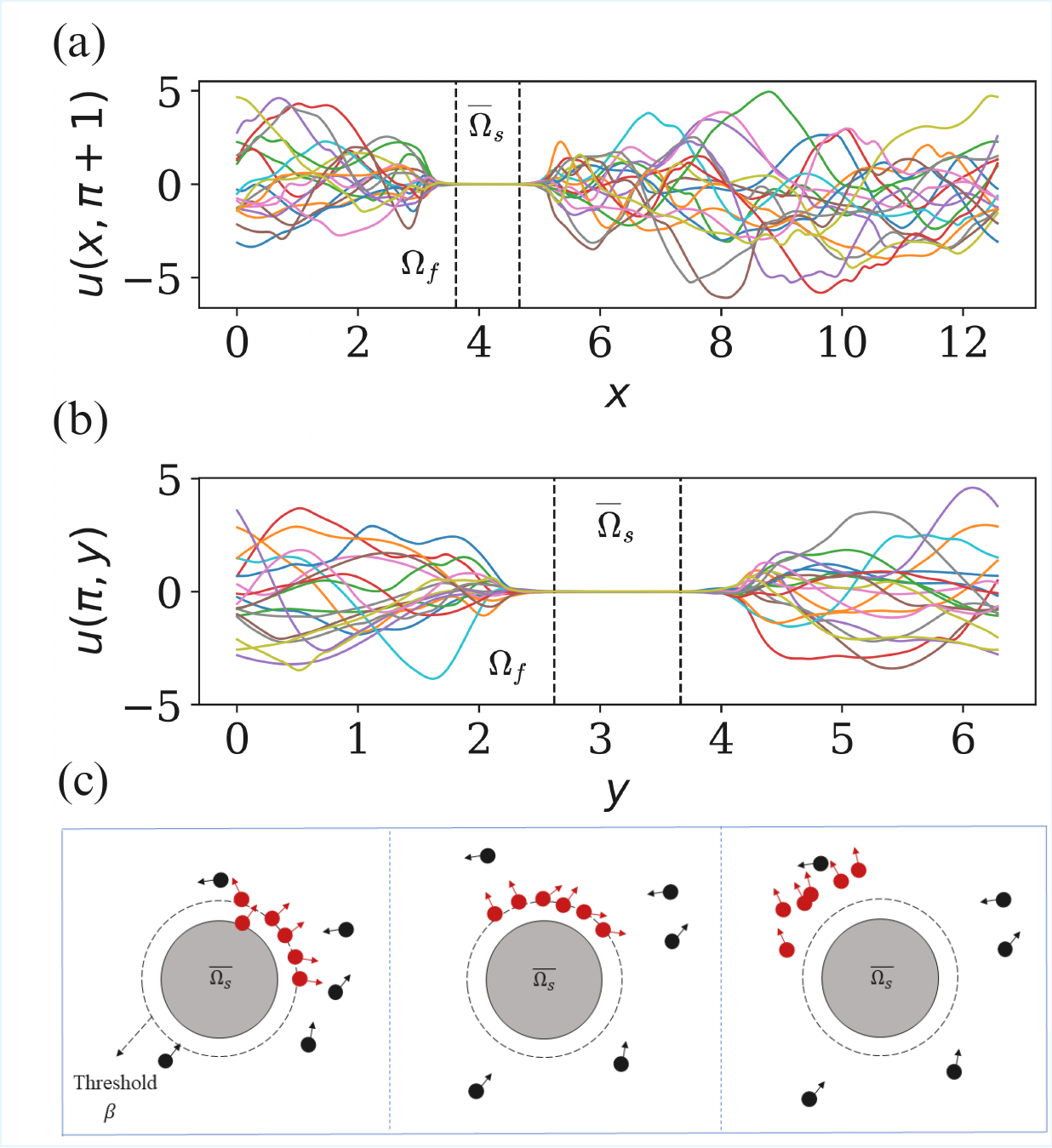}
    \caption{(a) - (b) \textbf{Plots showing the fluid velocity along lines intersecting the obstacle.} Vertical dashed lines indicate the obstacle boundaries.(a) Velocity profile along the line $y = \pi + 1$. The velocity of fluid goes to vanishes where the obstacle is present. (b) Velocity profile along the line $x = \pi$.  In both plots, different colors represent the velocity field $\bm {u}$ sampled every 10,000 time steps. 
    (c) \textbf{Schematic diagram of microswimmer interaction with the obstacle.} Red dots mark swimmers near the obstacle. This illustrates typical reorientation and detachment behavior, consistent with the results below.}
    \label{fig:Penalization}
\end{figure}
\newline
\noindent   

The remaining part of this paper is organised as follows. Section~\ref{sec:level1} presents our model and discusses the methods we use to study it. Section~\ref{sec:result} is devoted to a detailed analysis of our results. Section~\ref{sec:summary} concludes the paper by summarizing our key findings and their physical implications.

\section{\label{sec:level1}Model and Methods}

In \cref{subsec:fluid} we describe the incompressible Navier-Stokes (NS) equation for the background fluid, the volume-penalization method we use, and the pseudospectral method we employ. \cref{subsec:micro} presents our model for microswimmers and the numerical methods we use to study their dynamics. In \cref{subsec:machine_learning}, we describe our reinforcement-learning algorithms.

\subsection{Background fluid}
\label{subsec:fluid}

\noindent
The flow velocity of the background fluid is calculated using a 2D incompressible Navier-Stokes equation \cite{Pandit_2017, Lesieur_2008, Frisch_1995, PhysRevE.100.053101,Mukherjee_2019} with segregated solid and fluid domains. In particular, we employ the volume-penalization method, which modifies the Navier-Stokes equation to accommodate the boundary conditions and the permeability of the solid domain \cite{Schneider_2011, Kevlahan_2001, Kolomenskiy_2009,padhan2025cahn}.
\begin{eqnarray}
    \partial_t \omega + (\bm{u} \cdot \nabla)\omega &=& \nu \nabla^2 \omega -\nabla \times \left[\frac{\chi}{\eta}\bm{u} \right] - \alpha \omega +  F_{\omega}\,;\nonumber \\
    \nabla \cdot \bm{u} &=& 0\,.
    \label{eq:NS}
\end{eqnarray} 
\newline
\noindent
Here, $\bm{\omega} =  \nabla \times \bm{u} = \omega \hat{\bm{z}}$ is the vorticity and $\bm{u} \equiv (u_x, u_y)$ denotes the fluid velocity, $\alpha$ is the coefficient of friction, $\nu$ is the kinematic viscosity, and $F_{\omega}$ is the large-scale stochastic force that we apply to maintain statistically steady turbulence. The permeability of the obstacle is $\eta$ and $\chi$ is the mask function, which scales the velocity field throughout the domain and penalizes it by a factor of $\frac{1}{\eta}$. We implement volume-penalization in the following way \cite{Padhan_2024}:
\begin{align}
    \label{eq:tanh}
    \chi_\zeta &= 0.5\bigg( 1 + \tanh \bigg(\frac{R_0 - \sqrt{(x - x_0)^2 + (y-y_0)^2}}{\zeta}\bigg) \bigg)\,; 
\end{align}
$R_0$ is the radius of a circular obstacle, whose center lies at  $(x_0,y_0)$,
and which is separated from the rest of the domain by an interface of width $\zeta$. In the limit $\zeta \to 0$, the hyperbolic tangent function in Eq.~\eqref{eq:tanh} separates the domain into two areas distinguished by a sharp interface with the following map function~\cite{Kolomenskiy_2009} [see \Cref{fig:Penalization}]:
\begin{align}   
{\chi_{\zeta=0}} &= \begin{cases}    1, &  x \in \overline{\Omega_s} \\
    0,    &  x \in \Omega_f \,.
\end{cases}
\label{eq:chi}
\end{align}

In \Cref{fig:Penalization}(a) we depict the fluid velocity profile along the horizontal line $y = \pi + 1$; the fluid velocity vanishes inside the obstacle region (the vertical dashed lines indicate the spatial extent of the obstacle). Similarly, \Cref{fig:Penalization}(b) shows the velocity profiles along the vertical line $x = \pi$. In both plots, different colors represent the velocity field $\bm {u}$ sampled at every $10^4$ time steps; we use the parameters in Table~\ref{tab:parameters}. In \Cref{fig:Penalization}(c) we illustrate, schematically, how microswimmers can interact with the obstacle.

\begin{table}[h!]
\begin{tabular}{|l |c|r|}
\hline
$\alpha$  & Friction coefficient  & $5 \times 10^{-2}$ \\
$\nu$     & Kinematic viscosity   & $2 \times 10^{-3}$  \\
$\eta$    & Permeability          &$2 \times 10^{-4}$  \\
$u_{\mathrm{rms}}$ & RMS velocity  & 1.8  \\
$\omega_{\mathrm{rms}}$ & RMS vorticity     & 4.0 \\
$\widetilde{V_s}$  & Non-dimensional swimmer velocity & 0.834  \\
$\beta$               & Reflective constant                       & $5 \times 10^{-4}$                \\
$B$                   & Time scale for reorientation of direction & 0.1                     \\
$\lambda$             & Learning parameter                        & $10^{-2}$                  \\
$\gamma$              & Discount factor                           & 0.99                  \\
$\epsilon_g$          & Epsilon-greedy parameter                  & $10^{-3}$    \\ \hline
\end{tabular}
\caption{Table of parameters; $u_{\mathrm{rms}} \equiv \langle |u|^2 \rangle$, $\omega_{\mathrm{rms}} \equiv \langle |\omega|^2 \rangle$, and $\widetilde{V_s} \equiv \frac{V_s}{u_{\mathrm{rms}}}$;  RMS stands for root mean squared.
We use $\tau_\Omega \equiv \omega_{\mathrm{rms}}^{-1}$ to nondimensionalize time.}
 \label{tab:parameters}
\end{table}

To solve Eq.~\eqref{eq:NS} we use a pseudospectral direct numerical simulation (DNS) with periodic boundary coditions~\cite{Hussaini_1987,Pandit_2017,Tadmor_1991,verma_2020} in a rectangular domain of size $4\pi \times 2\pi$ that has $512 \times 256$ collocation points; we use the standard $2/3$ dealiasing scheme to remove aliasing errors~\cite{Hussaini_1987,biferale2019zermelo,Alageshan_2020}.
For time marching we employ the third-order exponential Runge-Kutta scheme \cite{Cox2002}. The values of all parameters are listed in Table~\ref{tab:parameters}. 

\begin{figure*}[!]
    \centering
    \includegraphics[width = 0.85\textwidth]{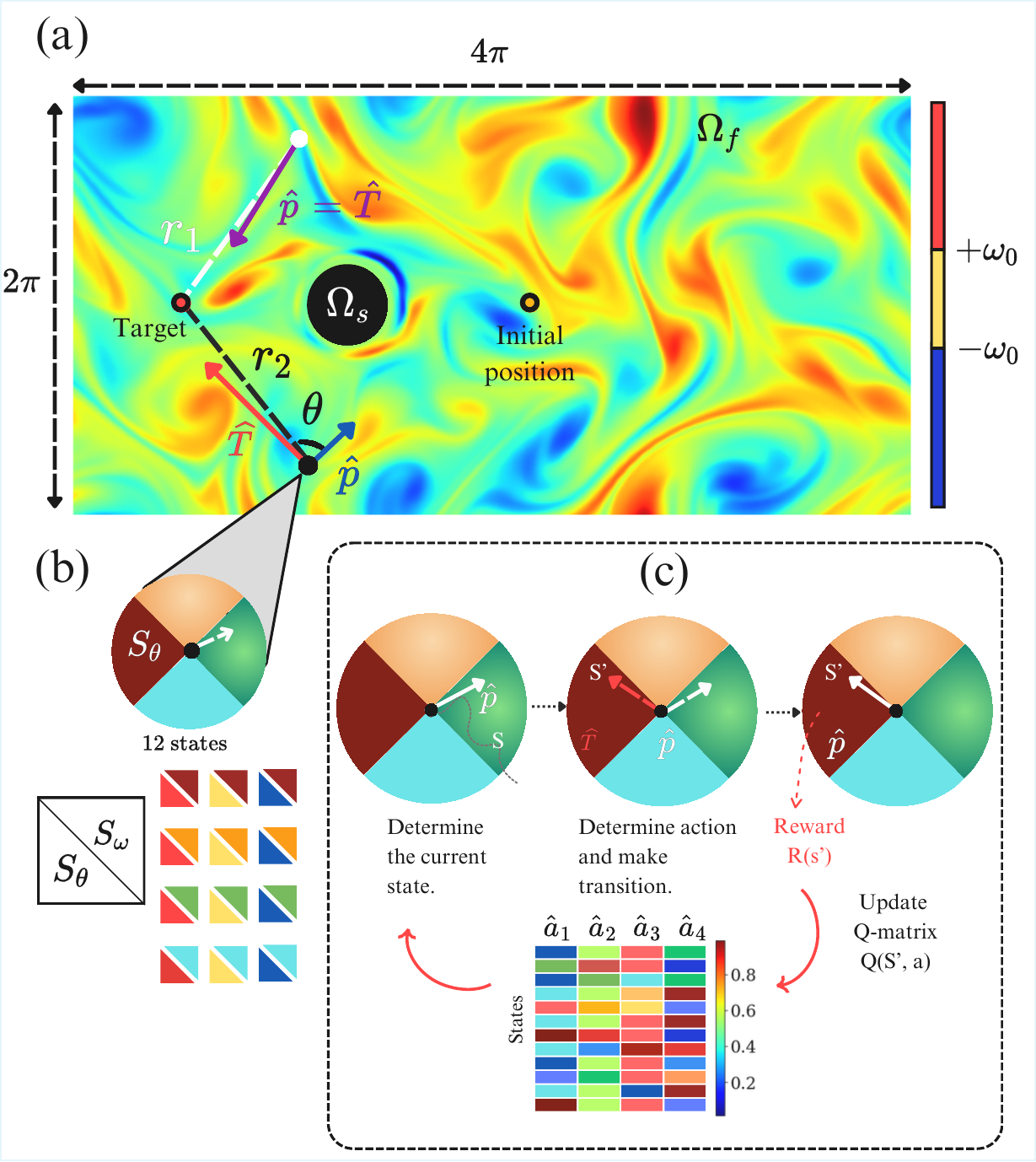}
    \caption{ \textbf{(a)} provides a visual snapshot of microswimmers within a turbulent fluid. Here, $\Omega_s$ represents the solid, volume-penalized domain, while $\Omega_f$ denotes the fluid domain. The root mean squared vorticity, $\omega_0$, serves as a threshold used to determine the state of the microswimmer. The na\"ive swimmer is depicted as a white dot, while the smart swimmer is shown as a black dot. The distances between the na\"ive and smart swimmers and the target are denoted by $r_1$ (white dotted line) and $r_2$ (black dotted line), respectively. The symbol $\hat{T}$ denotes the swimmer’s direction relative to the target and acts as the control direction. The na\"ive swimmer constantly reorients itself to point directly toward the target and moves along that line, with the purple arrow indicating this orientation, as opposed to to a smart swimmer that learns to orient itself to the target through the reinforcement process. 
    The target is positioned at $(\tilde{x}, \tilde{y}) = (0.1704, 0.5)$, while the swimmers begin at $(\tilde{x}, \tilde{y}) = (0.5795, 0.5)$. The coordinates $(\tilde{x}, \tilde{y})$ are non-dimensionalized positions as $(x/L_x, y/L_y)$ respectively.
    This specific arrangement was selected to create a scenario where the distance between the starting point and the target is shorter when navigating through the obstacle, but longer if the swimmers attempt to bypass the boundary. This setup provides an opportunity to observe the path-planning strategies employed by the smart swimmers.
    \textbf{(b)} shows the process of state construction. The current state is defined based on the combination of two main parameters: the fluid vorticity (three states, indicated by the red, yellow, and blue colorbar in subfigure (a)) and the angle relative to the target (denoted as $S_\theta$ represented by dark red, light blue, orange, and green). This approach results in a total of 12 possible states.
    \textbf{(c)} depicts the Q-learning process as stated in \ref{machine_learning}.  
    \label{fig:ML_bg_flow} }
\end{figure*}
\subsection{Microswimmers}
\label{subsec:micro}
\noindent
We consider $10^4$ 
noninteracting microswimmers.  The orientation of a microswimmer is represented by $\hat{\bm{p}}$.  We have two types of microswimmers in our simulation domain:  na\"ive swimmers (NS) and smart microswimmers (SS). The former determine their direction as $\widehat{\bm{T_i}} = {\bm{X^T} - \bm{X_i^{\text{NS}}}}/{|\bm{X^T} - \bm{X_i^{\text{NS}}}|}$, where $\bm{X}_i^{\text{NS}}$ denotes the position of the $i^{th}$ na\"ive swimmer and $\bm{X}^T$ is the target position, so that $\widehat{\bm{T_i}}$ is the unit vector pointing from the swimmer towards the target. Smart Swimmers (SS) utilize reinforcement machine learning and regulate their actions at every iteration according to environmental queues. 

A microswimmer's velocity is $\bm{V_s} = V_s\hat{\bm{p}}$; and its position $\bm{X}$ evolves as follows~\cite{pedley1992hydrodynamic,Alageshan_2020}:
\begin{align}
\frac{d\bm{X}}{dt} = \bm{u}(\bm{X}, t) + V_s\bm{\hat{p}}\,,\label{dX_dt}
\end{align}
where $u(\bm{X}, t)$ is the fluid velocity at $\bm{X}$. Thus, a microswimmer is propelled not only by the background flow, but also because of its own velocity $\bm{V_s}$; we choose $V_s\simeq u_{rms}$. 
The orientation of the microswimmer is given by~\cite{pedley1992hydrodynamic}
\begin{eqnarray}
	\frac{d\hat{\bm{p}}}{dt} &=& \frac{1}{2B} \left[\hat{\bm{o}}-(\hat{\bm{o}}
	     . \hat{\bm{p}}) \hat{\bm{p}} \right] + \frac{1}{2} \bm{\omega} \times
	      \hat{\bm{p}}\,,
    \label{dP_dt}
\end{eqnarray}
where $\hat{\bm{o}}$, the control direction, points from $\bm{X}$ towards the target for a \textit{na\"ive} microswimmer. Note that $\hat{\bm{p}}$ tries to align with $\hat{\bm{o}}$, and $B$ is the time scale associated with this alignment. We account for the obstacle by incorporating volume penalization, via the mask function Eq.~\eqref{eq:chi}, and reformulating Eq.~\eqref{dX_dt} as follows:
\begin{align}
\frac{d\bm{X}}{dt} = \bm{u}(\bm{X}, t) + \bm{\hat{p}} \left[V_s  - \frac{V_e}{\beta}\chi_\zeta\right]\,,  \label{eq:dX_dt} 
\end{align}
where $V_e$ is the escape velocity and $\beta$ the reflectivity constant taken for the volume-penalized domain. As a swimmer approaches the obstacle region ($\chi_\zeta \to 1$), its effective propulsion speed is reduced and can reverse sign if $V_e/\beta > V_s$. This phenomenological rule discourages sustained penetration into the obstacle and biases the swimmer motion away from the penalized region. We emphasize that it does not impose a strict geometric reflection and does not guarantee escape for all instantaneous orientations.

 We also compare the performance of our smart swimmer with surfers (SuS), which use local hydrodynamics and velocity gradients for navigation and are known for adapting to evolving turbulent environments, especially for the upward direction in phototaxis [as studied in three dimensions in \cite{monthiller2022surfing}]. Surfers follow Eqs.~\eqref{dP_dt} and ~\eqref{eq:dX_dt}, but with the control direction given by
\begin{equation}
    \bm{\widehat{o}} = \widetilde{\exp \left( \tau \nabla \bm{u} \right)}\cdot \bm{\widehat{T}}\,,
\end{equation}
\noindent
where the tilde denotes a transpose, $\widehat{\bm{T}}$ is the direction towards the target, and the time horizon $\tau$ is a free parameter in the surfing strategy.


\subsection{\textbf{Machine learning and adversarial $\mathcal{Q}$ learning}}  \label{machine_learning}
\label{subsec:machine_learning}
\noindent
We develop a reinforcement-learning strategy that optimizes the path planning of a microswimmer which tries (a) to reach a target, while navigating through a turbulent flow, and (b) to avoid stalling in stagnation regimes in the vicinity of an obstacle. 
For part (a) we use the adversarial-reinforcement-learning framework developed \cite{Alageshan_2020}. In particular, we use an adversarial-reinforcement-learning scheme [see the flowchart in \Cref{flowchart}], in which each microswimmer, a master, is accompanied by a \textit{slave swimmer} (SLS)~\cite{Kong2017RevisitingTM} that follows the  na\"ive strategy of pointing towards the target; the position of SLS is denoted by $\bm{X^{\text{SLS}}}$. 

For a practical reinforcement-learning  scheme, we discretize $\omega$ and 
$\theta\equiv \cos^{-1}\big(\hat{\bm{T}}\cdot\hat{\bm{p}}\big)$, where $\hat{\bm{T}}\equiv(\bm{X}^{T}-\bm{X}(t))/\vert\bm{X}^{T}-\bm{X}(t)\vert$, for a given the swimmer, is the unit vector that points from the microswimmer to the target [where $\bm{X}^{T}$ and $\bm{X}$ are the position vectors of the fixed target and the microswimmer, respectively]. $\omega$ assumes the following $3$ states $S_\omega$: 
\begin{eqnarray}
\omega_1,\, &\rm{if}&\, \omega>\omega_{0}\,;\nonumber \\
\omega_2,\, &\rm{if}&\, \omega_{0} \le \omega \le \omega_{0}\,;\nonumber \\
\omega_3,\, &\rm{if}&\,  \omega < -\omega_{0}\,;
\end{eqnarray}
$\omega_{0}=0.5 \omega_{rms}$, where the subscript $rms$ denotes root-mean-squared [see the colorbar in ~\Cref{fig:ML_bg_flow} (a)]. $\theta$ assumes the following $4$ states $S_\theta$ [see the circle in \Cref{fig:ML_bg_flow} (b)]:
 \begin{eqnarray}
\theta_1,\, &\rm{if}&\, -\pi/4 \leq \theta < \pi/4 \,; \nonumber \\
\theta_2,\, &\rm{if}&\,  \pi/4\leq \theta < 3\pi/4 \,; \nonumber  \\
\theta_3,\, &\rm{if}&\,  3\pi/4\leq \theta <5\pi/4 \,; \nonumber \\
\theta_4,\, &\rm{if}&\, -3\pi/4 \leq \theta < -\pi/4\,.
 \end{eqnarray}
These discrete states are illustrated in \Cref{fig:ML_bg_flow} which consists of several schematic diagrams. The first of these, \Cref{fig:ML_bg_flow} (a), 
is a representative pseudocolor plot of the vorticity $\omega$ of the turbulent fluid with the instantaneous positions of two microswimmers -- one na\"ive and the other smart -- superimposed on it; $\Omega_s$ is the solid, volume-penalized domain and $\Omega_f$ denotes the fluid domain. The na\"ive swimmer (NS) is depicted by a white filled circle, whereas the smart swimmer (SS) is shown via a black-filled circle; $r_1$ and $r_2$ are, respectively, the distances between the target and NS and SS. For the NS, $\hat{\bm{p}} = \hat{\bm{T}}$ at all times $t$.
\Cref{fig:ML_bg_flow} (b) illustrates how $S_\omega$ and $S_\theta$ together yield $12$ states [see the 12 rows in \Cref{fig:ML_bg_flow}(c)] of the
set $\mathcal{S}$, the Cartesian product  $S_\omega \otimes S_\theta$, which label the rows of our $\mathcal{Q}$ matrix. 
The set of actions $\mathcal{A} \equiv \{\bm{A}_{1},\bm{A}_{2},\bm{A}_{3},\bm{A}_{4}\}\equiv\{\hat{\bm{T}},-\hat{\bm{T}},\hat{\bm{T}}_{\perp},-\hat{\bm{T}}_{\perp}\}$
label the columns of $\mathcal{Q}$  [see \Cref{fig:ML_bg_flow} (c)]. All possible discrete states of the microswimmers are denoted by
colored squares whose the lower halves stands for the vorticity state,
 $S_\theta$, and the upper halves represents the direction state, $S_\omega$.
\begin{figure*}[!]
    \centering
  \includegraphics[width = \linewidth]{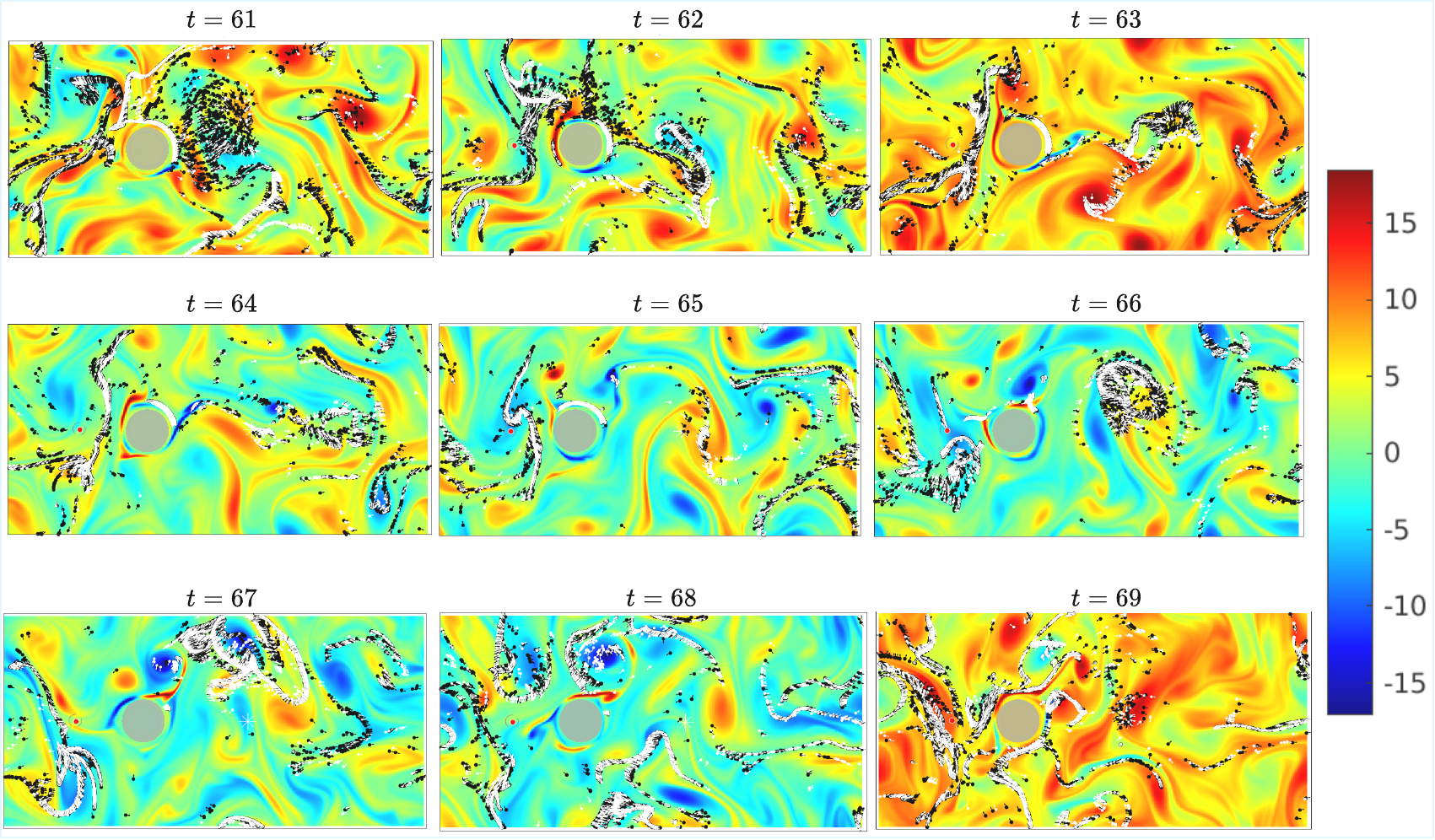}
    \caption{Pseudocolor plots of the vorticity $\omega$ with superimposed microswimmer positions. The arrows denote the directions of their velocities at nine representative times, showing the spatiotemporal evolution [see the Video V1 in the Supplemental Information] of the flow and the microswimmers. In the time interval $t = 61 ~\text{to}~ 63$, the swimmers interact with the obstacle and adjusting their alignment to move away from it. The sequence from $t=64 ~\text{to}~68$ illustrates how the swimmers glide along the obstacle, gradually realigning until they fully detach. In the snapshot at $t=69$, new microswimmers are shown attaching to the obstacle, thereby continuing the interaction cycle. }
    \label{vorticity_evo}
\end{figure*}
The $\mathcal{Q}$-learning method updates the value of $\mathcal{Q}$, which estimates the expected optimal return for any smart microswimmer (SS) in a given state $\bm{s}(t)\in \mathcal{S}$, using the Bellman equation \cite{Sutton_1988}, whenever a state transition occurs:
\begin{eqnarray}
    {\mathcal{Q}}(\bm{s}(t),\hat{\bm{o}}(\bm{s}(t))) &\mapsto& (1-\lambda) \, \mathcal{Q}(\bm{s}(t),\hat{\bm{o}}(\bm{s}(t))) \nonumber \\
    &+& \lambda \, \left[ \mathcal{R}(t) + \gamma \max_{\hat{\bm{a}}} \mathcal{Q}(\bm{s}(t+\Delta t),\hat{\bm{a}}) \right]\,; \nonumber \\
    \mathcal{R}(t)\equiv|\bm{X^{\text{SLS}}}(t) &-& \bm{X^{{T}}}|-|\bm{X^{\text{SS}}}(t)-\bm{X^{{T}}}|\,.
    \label{Q_eqn}
\end{eqnarray}
Here, \( \lambda \) is the learning rate,  \( \mathcal{R}(t) \) the reward, \( \gamma \) the discount factor emphasizing the significance of future rewards, the action $\hat{\bm{a}} \in \mathcal{A}$, and the superscripts SLS, SS, and $\bm{T}$ denote
the slave swimmer, smart swimmer, and target, respectively.
We optimise the strategy of the microswimmer by choosing the best action as follows:
\begin{equation}
    \hat{\bm{o}}(\bm{s}(t)) = \arg\max_{\hat{\bm{a}}}\{ \mathcal{Q}(\bm{s}(t),\hat{\bm{a}}) \}\,.
\end{equation}
\noindent
 In \cref{fig:ML_bg_flow}(c), we illustrate the scheme for calculating the reward function of the smart microswimmer. The orientation of the slave microswimmers, in state $S(\omega, \theta)$  is indicated by the white arrow. In the subsequent figure, the smart microswimmer changes its state to $S'(\omega, \theta)$  (red arrow)  after selecting a new action from the $\mathcal{Q}$-matrix via $arg\max_{a_i}\{ \mathcal{Q}(s_i,a_i) \}$. After computing the reward function, $R(t) = |\bm{X^{\text{SLS}}_i} - \bm{X^{{T}}}| - |\bm{X^{\text{SS}}_i} - \bm{X^{{T}}}|$ \cite{Alageshan_2020,gupta2024flockingaidpathplanning}, the $\mathcal{Q}$-matrix is updated from  $\mathcal{Q}(s,a)$ to $\mathcal{Q}(s',a)$. The flow chart in \Cref{flowchart} illustrates the sequence of steps involved in our adversarial $\mathcal{Q}$-learning framework for smart microswimmers. 

\noindent

\section{Results}

\label{sec:result}
\begin{figure}[!]
    \centering
    \includegraphics[width=0.8\linewidth]{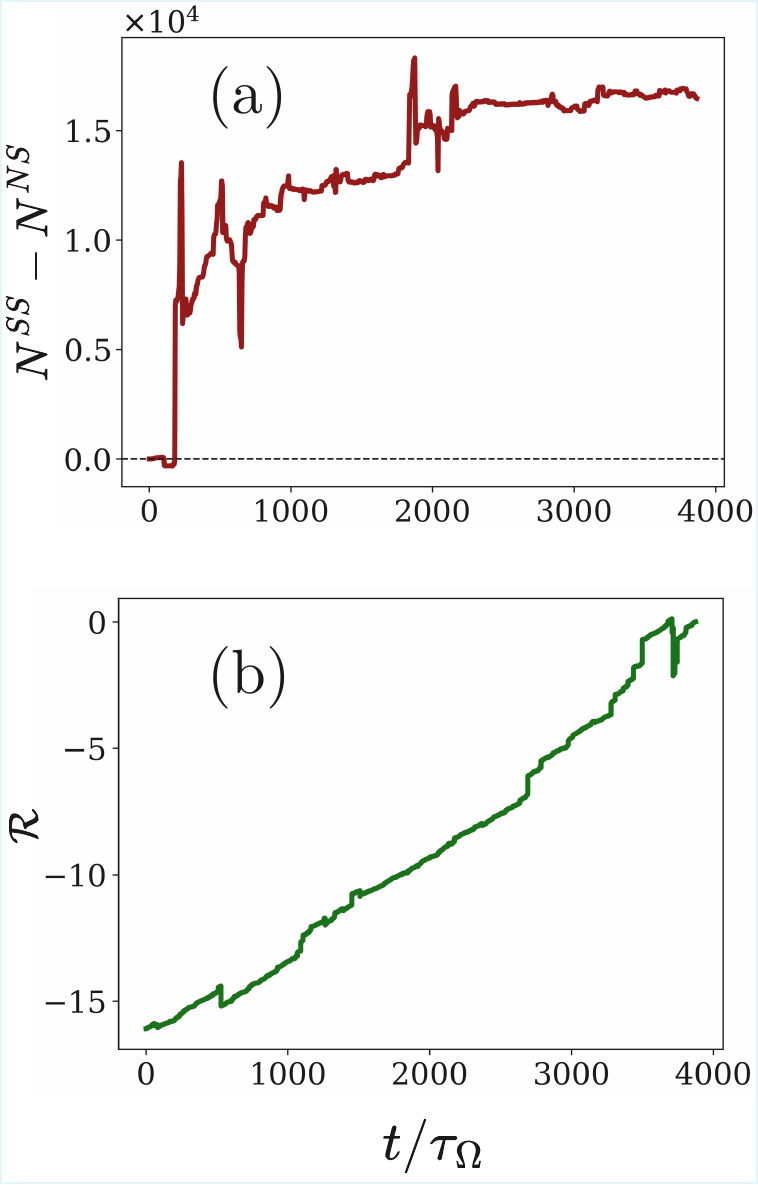}
    \caption{ Plots vs the nondimensionalized time $t/\tau_{\Omega}$; (a) Cumulative sum of the difference between na\"ive and smart microswimmers reaching the target; (b) the cumulative sum of the rewards obtained.}
    \label{Results}
\end{figure}

\begin{figure}
    \centering
    \includegraphics[width=\linewidth]{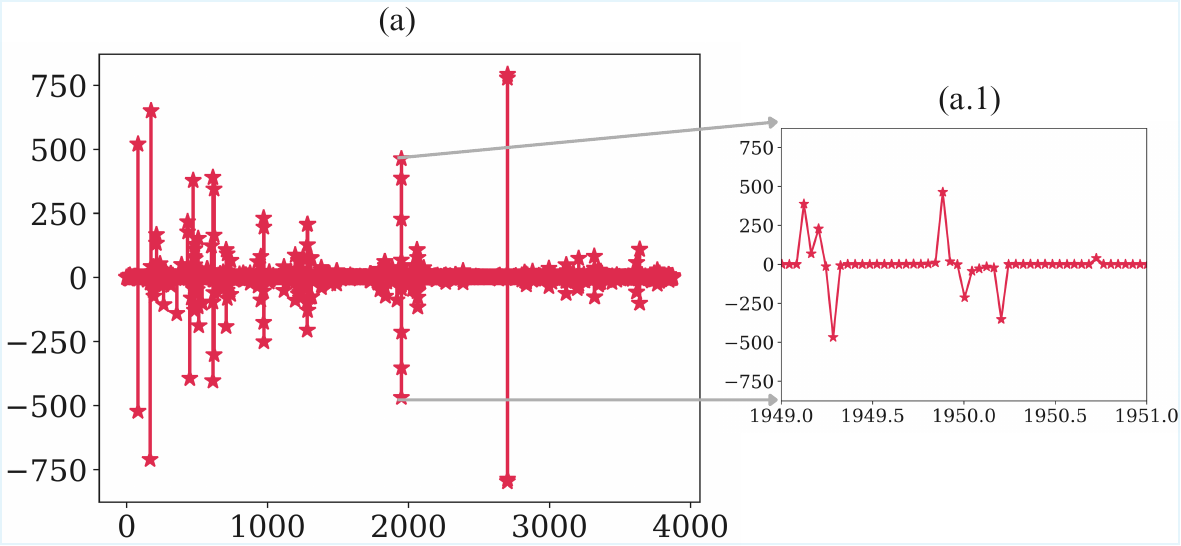}
    \caption{Plot (a) displays the difference in Smart and na\"ive microswimmers interacting with the obstacle $\Delta$, at the detachment interval $\delta t$ of 1000 steps; (a.1) a zoomed version in the vicinity of one of the spikes.}
    \label{stuck}
\end{figure}

\begin{figure}
    \centering
    \includegraphics[width=1.1\linewidth]{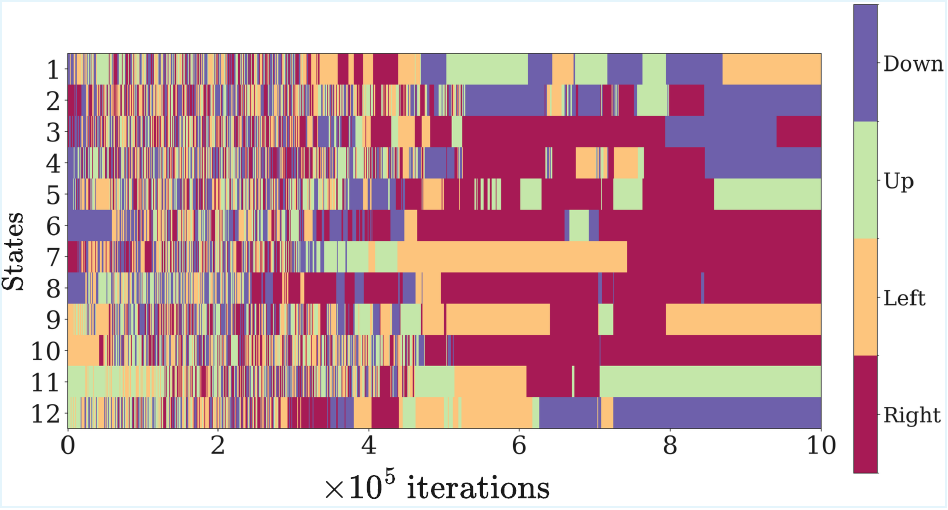}
    \caption{ This plot shows the time series of the maximal value in a row of the $\mathcal{Q}$ matrix at each for all $12$ states. These values correspond to the action set $\mathcal{A} := \big\{\widehat{\bm{T}}, -\widehat{\bm{T}}, \widehat{\bm{T}_\perp}, -\widehat{\bm{T}_\perp} \big\}$, which we interpret as the directions \textit{[Left, Right, Up, Down]} relative to the target; after the $\mathcal{Q}$-matrix stops evolving, then this plot does not show changes from row to row. }
    \label{Qseries_TL}
\end{figure}

\begin{figure*}[!]
 
    \centering
    \includegraphics[width = 0.8\linewidth]{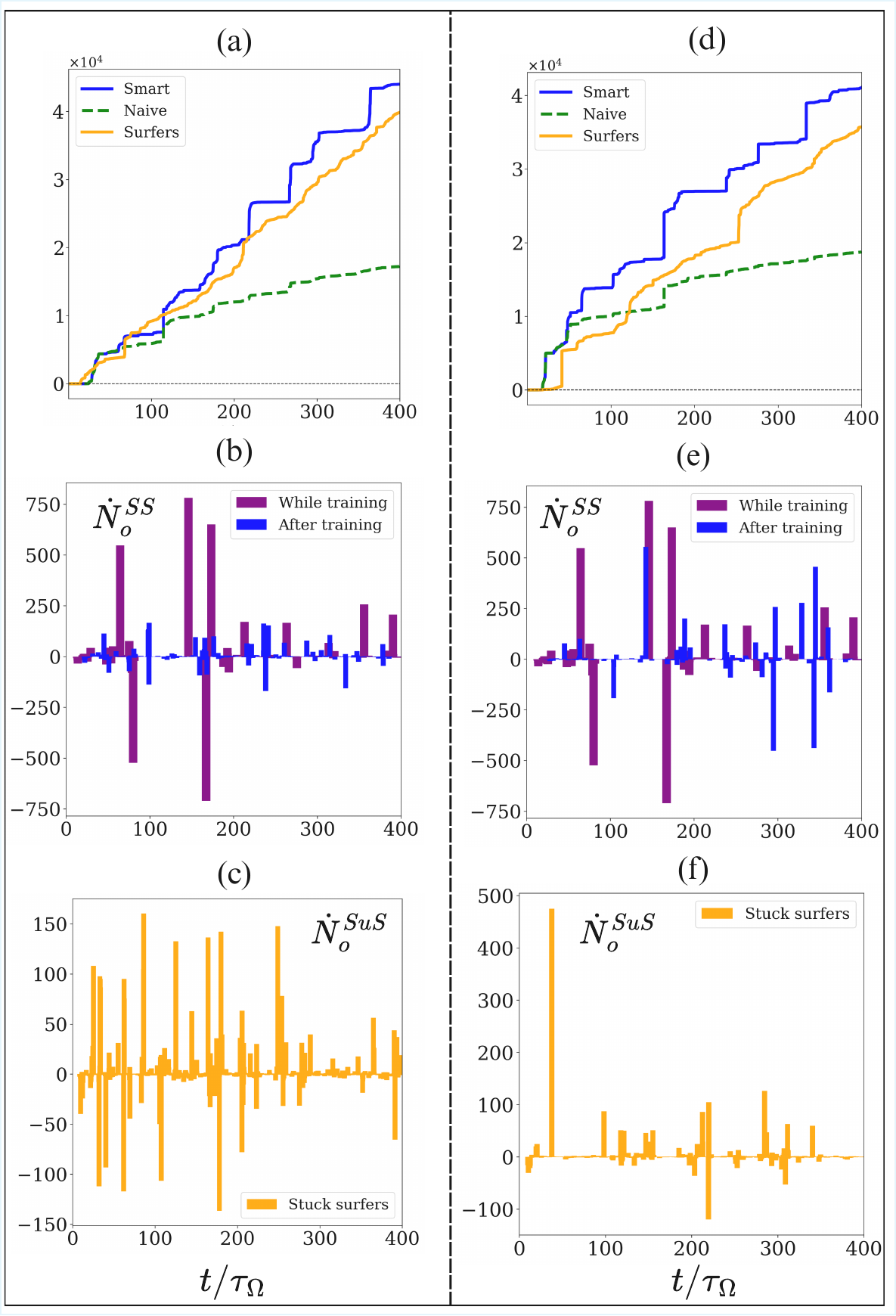}
\caption{\textbf{Plots (a), (b), (c)} depict the performance of the swimmers when they are initialized in the same position  $(\tilde{x}, \tilde{y}) = (0.5795, 0.5)$ as their training: (a) we plot total numbers of smart swimmers $N^{SS}(t)$, na\"ive swimmers $N^{NS}(t)$, and surfers $N^{SuS}(t)$ reaching the target vs time; (b) compares the number of smart swimmers interacting with the obstacle during the training and  after the training; (c) the number of surfing swimmers interacting with the obstacle. \textbf{Plot (d), (e), (f)} depict the performance of the swimmers when they are initialized at a different position than the training. i.e., \((\tilde{x}, \tilde{y}) = (0.5795, 0.65)\): (d) the total numbers of smart swimmers $N^{SS}(t)$, na\"ive swimmers $N^{NS}(t)$, and surfers $N^{SuS}(t)$ reaching the target vs time; (e) comparison of the number of smart swimmers interacting with the obstacle during the training and  after the training;
(f) the number of surfing swimmers interacting with the obstacle.    
\label{Fig7}}
\end{figure*}

\begin{figure}[!]
    \centering
    \includegraphics[width=\linewidth]{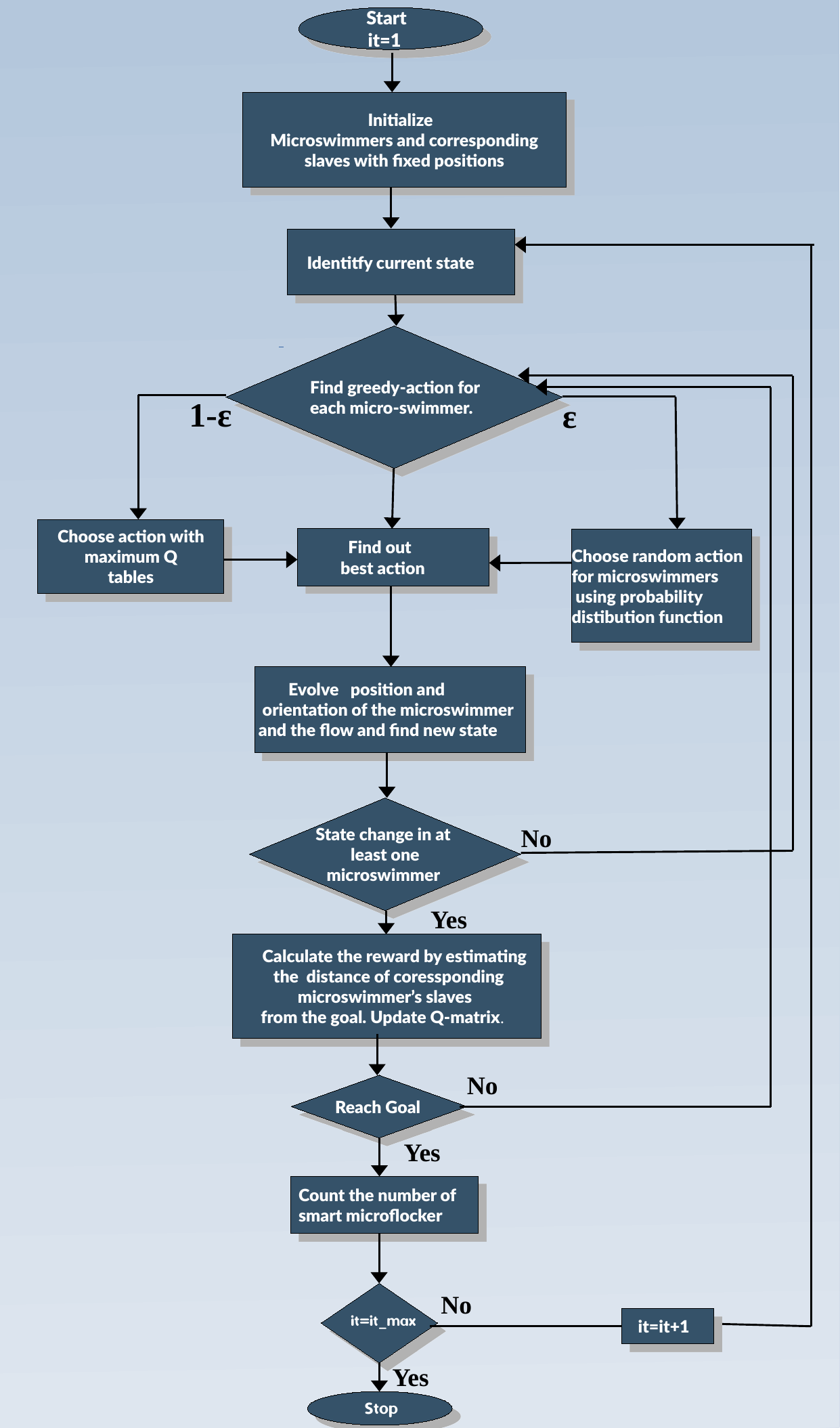}
    \caption{Flow chart of sequence of processes involved in adversarial Q learning scheme. }
    \label{flowchart}
\end{figure}

\noindent We present the principal results of our study in \crefrange{vorticity_evo}{Fig7}.
\newline 
\noindent 
\Cref{vorticity_evo} presents pseudocolor plots of the vorticity field  with superimposed microswimmer positions, where the arrows denote the instantaneous directions of their velocities. These plots, shown at nine representative times, capture the spatiotemporal evolution of the flow field together with the collective dynamics of the microswimmers. 
During the interval $t=61-63$, the microswimmers encounter the obstacle and begin to interact with it. This interaction is reflected in their reorientation, as they adjust their swimming directions to navigate away from the immediate vicinity of the obstacle. In the subsequent frames ($t=64-68$), the swimmers exhibit a gliding motion along the obstacle’s surface and a gradual realignment of their swimming orientation until they eventually detach completely from the boundary.
At $t=69$, a new set of swimmers  approaches the obstacle, thereby initiating a fresh round of interactions. This sequence highlights the cyclic nature of the swimmer–obstacle interactions: initial approach, alignment adjustment, gliding and reorientation, detachment, and eventual replacement by new swimmers. 

\noindent

We first define $N^{SS}(t)$, $N^{NS}(t)$, and $N^{SuS}(t)$ which are, respectively, the total numbers of smart microswimmers, naïve swimmers, and surfers that have reached the target up until $t/\tau_\Omega$ for the parameters in Table~\ref{tab:parameters}.
The performance of smart swimmers compared to na\"ive ones is analyzed conveniently by plotting  $N^{SS}(t)-N^{NS}(t)$ versus 
$t/\tau_\Omega$ as shown in \Cref{Results} (a); this demonstrates that smart microswimmers perform better than na\"ive ones rapidly. 
In \Cref{Results} (b) we plot the reward $\mathcal{R}(t)$ as a function of $t/\tau_\Omega$; the upward trend of this curve signifies that smart swimmers are increasingly making effective decisions. 

Figure~\ref{stuck} displays the difference between smart and na\"ive microswimmers that interact with the obstacle; here, 
\( N_o^{SS} \) and \( N_o^{NS} \) denote, respectively, the numbers of smart and na\"ive swimmers. 
Over a time interval \( \delta t \), the rates of change of these numbers are:
\begin{eqnarray}
    \dot{N}^{SS}_o &=& \frac{N^{SS}_o(t) - N^{SS}_o(t + \delta t)}{\delta t}\,; \nonumber \\
    \dot{N}^{NS}_o &=& \frac{N^{NS}_o(t) - N^{NS}_o(t + \delta t)}{\delta t}\,.
    \label{eq:Ndot}
\end{eqnarray}
Positive values of these rates indicate net detachment from the obstacle;  
the differential detachment rate
\begin{equation}
    \Delta \equiv \dot{N}^{SS}_o - \dot{N}^{NS}_o
\end{equation}
is positive whenever smart swimmers detach, from the obstacle, more efficiently than na\"ive swimmers.   
The obstacle obstructs the shortest path, from the microswimmer to the target, so repeated interactions occur between these swimmers and the obstacle; these interactions lead to sharp peaks or troughs in the plot of $\Delta$ versus time $t$ in \Cref{stuck}(a), a zoomed version of which is given in \Cref{stuck}(a.1).

Figure~\ref{Qseries_TL} shows the maximal value in a row of the $\mathcal{Q}$ matrix versus time, for each one of the $12$ states [\Cref{fig:ML_bg_flow}]; we employ the action set $\mathcal{A} := \big\{\widehat{\bm{T}}, -\widehat{\bm{T}}, \widehat{\bm{T}_\perp}, -\widehat{\bm{T}_\perp} \big\}$, which we interpret as the directions \textit{[Left, Right, Up, Down]} relative to the target. This plot does not show changes, from row to row, after the $\mathcal{Q}$-matrix stops evolving.
After optimizing the $\mathcal{Q}$ matrix, we check the performance of smart microswimmers by choosing the best action from the optimized $\mathcal{Q}$ matrix; we find that smart-microswimmer performance surpasses that of na\"ive microsswimmers. 


\noindent
\noindent 
The number of surfing swimmers interacting with the obstacle is $N^{SuS}_o(t)$; over a time interval \( \delta t \), the rates of change of surfing swimmers are [cf. Eq.~\eqref{eq:Ndot}]
\begin{eqnarray}
    \dot{N}^{SuS}_o &=& \frac{N^{SuS}_o(t) - N^{SuS}_o(t + \delta t)}{\delta t}\,.
    \label{eq:Nsurfdot}
\end{eqnarray}
In \Cref{Fig7} we present plots of $N^{SS}(t)$, $N^{NS}(t)$, and $N^{SuS}(t)$ and related quantities. The plots in the first column, namely, \Cref{Fig7} (a), (b), and (c) are obtained when the swimmers start from the initial position  $(\tilde{x}, \tilde{y}) = (0.5795, 0.5)$, shown in \Cref{fig:ML_bg_flow} (a). The plots in the second column, \Cref{Fig7} (d), (e), and (f) are obtained when the swimmers start from the position $(\tilde{x}, \tilde{y}) = (0.5795, 0.65)$ mentioned in the caption of \Cref{Fig7}. Figures~\ref{Fig7} (a) and (d) show clearly that smart swimmers outperform both surfers and na\"ive swimmers. Figures~\ref{Fig7} (b) and (e) present plots of  $\dot{N}^{SS}_o$ during (purple) and after (blue) training; in the latter case we use the fixed $\mathcal{Q}$ matrix that we obtain after the training is over. Figures~\ref{Fig7} (c) and (f) present plots of  $\dot{N}^{SuS}_o$.

\section{Summary and discussion}
\label{sec:summary}

We have  introduced an obstacle-aware navigation framework for microswimmers in turbulent flows by incorporating volume penalization into a forced 2D Navier–Stokes environment. This approach enables microswimmers to sense nearby obstacles and escape from hydrodynamic stagnation regions in the vicinity of the obstacle. To improve navigation efficiency, our study incorporates an adversarial $\mathcal{Q}$-learning framework that enables microswimmers to learn from their previous interactions with the ambient flow and nearby obstacles. Through iterative updates of the $\mathcal{Q}$ matrix, the swimmers develop an effective form of memory, which allows them to refine their decision-making and progressively optimize their paths in their journeys towards the target. The performance of these swimmers is assessed using four key measures: a higher target-reaching rate relative to naïve swimmers; a successful escape from obstacle-induced trapping; a steady increase in cumulative rewards during training; and eventual convergence reflected in the stabilization of the $\mathcal{Q}$ matrix. Our results show that the trained smart microswimmers significantly outperform both naïve swimmers and surfer-type agents; and they maintaining this superior navigation efficiency even when deployed from different initial positions in similar flow environments. Our results have obvious physical applications: in particular, we provide an obstacle-aware navigation framework that offers a viable approach for studying and guiding microswimmers in turbulent flows with embedded obstacles~\cite{Takagi2013HydrodynamicCO, Tan2024}.

\section*{Data and code availability}
\noindent
Data from this study and the computer scripts can be obtained from the authors upon reasonable request.

\section*{Conflicts of Interest}
No conflicts of interest, financial or otherwise, are declared by the authors.
\section*{Author Contributions} 
\noindent 
VG, RP, and AG planned the research; VG, AG carried out the calculations and analysed the numerical data; VG prepared the tables, figures, and the draft of the manuscript; NBP contributed to the numerical methods part of the work; VG, AG, NBP, and RP then revised the manuscript in detail and approved the final version.
\section*{Acknowledgments}
\noindent
We thank K.V. Sai Swetha, K.V. Kiran, J.K. Alageshan, and V.K. Babu for the valuable discussions, the Anusandhan National Research Foundation (ANRF), the Science and Engineering Research Board (SERB), and the National Supercomputing Mission (NSM), India, for support,  and the Supercomputer Education and Research Centre (IISc), for computational resources. AG thanks DST-SERB for the NPDF fellowship (PDF/2021/001681).   
\clearpage
\section{Appendix}
\bibliography{bibfile}

\begin{thebibliography}{10}

\bibitem{Alageshan_2020}
Jaya~Kumar Alageshan, Akhilesh~Kumar Verma, J\'er\'emie Bec, and Rahul Pandit.
\newblock Machine learning strategies for path-planning microswimmers in turbulent flows.
\newblock {\em Phys. Rev. E}, 101:043110, Apr 2020.

\bibitem{alegado_king_2014_bacterial}
Rosanna~A. Alegado and Nicole King.
\newblock Bacterial influences on animal origins.
\newblock {\em Cold Spring Harbor Perspectives in Biology}, 6(11):a016162, 2014.

\bibitem{AlonsoMatilla2019}
Roberto Alonso-Matilla, Brato Chakrabarti, and David Saintillan.
\newblock Transport and dispersion of active particles in periodic porous media.
\newblock {\em Physical Review Fluids}, 4(4):043101, 2019.

\bibitem{Berg_2005}
Richard~M. Berry.
\newblock {E. coli} in motion.
\newblock {\em Physics Today}, 58(2):64--65, 02 2005.

\bibitem{biferale2019zermelo}
L.~Biferale, F.~Bonaccorso, M.~Buzzicotti, P.~Clark Di~Leoni, and K.~Gustavsson.
\newblock Zermelo's problem: Optimal point-to-point navigation in 2d turbulent flows using reinforcement learning.
\newblock {\em Chaos: An Interdisciplinary Journal of Nonlinear Science}, 29(10), October 2019.

\bibitem{boffetta2012two}
Guido Boffetta and Robert~E. Ecke.
\newblock Two-dimensional turbulence.
\newblock {\em Annual Review of Fluid Mechanics}, 44(1):427--451, 2012.

\bibitem{Calascibetta2022Taming}
Chiara Calascibetta, Luca Biferale, Francesco Borra, Antonio Celani, and Massimo Cencini.
\newblock Taming lagrangian chaos with multi-objective reinforcement learning.
\newblock {\em The European Physical Journal E}, 46(9), 2023.

\bibitem{Chakraborty2025SmartNavigation}
Priyam Chakraborty, Rahul Roy, and Shubhadeep Mandal.
\newblock Smart navigation of microswimmers in poiseuille flow via reinforcement learning.
\newblock {\em The European Physical Journal E}, 48:30, 2025.

\bibitem{Chepizhko2013}
Oleksandr Chepizhko and Fernando Peruani.
\newblock Diffusion, subdiffusion, and trapping of active particles in heterogeneous media.
\newblock {\em Physical Review Letters}, 111:160604, 2013.

\bibitem{Colabrese_2017}
Simona Colabrese, Kristian Gustavsson, Antonio Celani, and Luca Biferale.
\newblock Flow navigation by smart microswimmers via reinforcement learning.
\newblock {\em Phys. Rev. Lett.}, 118:158004, Apr 2017.

\bibitem{Colabrese_Gustavsson_Celani_Biferale_2018}
Simona Colabrese, Kristian Gustavsson, Antonio Celani, and Luca Biferale.
\newblock Smart inertial particles.
\newblock {\em Phys. Rev. Fluids}, 3:084301, Aug 2018.

\bibitem{Cox2002}
S.~M. Cox and P.~C. Matthews.
\newblock Exponential time differencing for stiff systems.
\newblock {\em Journal of Computational Physics}, 176(2):430--455, 2002.

\bibitem{Daddi-Moussa-Ider_2021}
Abdallah Daddi-Moussa-Ider, Hartmut Löwen, and Benno Liebchen.
\newblock Hydrodynamics can determine the optimal route for microswimmer navigation.
\newblock {\em Communications Physics}, 4, 2021.

\bibitem{Khiyati2023Steering}
Zakarya El~Khiyati, Raphaël Chesneaux, Laëtitia Giraldi, and Jérémie Bec.
\newblock Steering undulatory microswimmers using reinforcement learning.
\newblock {\em The European Physical Journal E}, 46(43), 2023.

\bibitem{Elgeti_2015}
J~Elgeti, R~G Winkler, and G~Gompper.
\newblock Physics of microswimmers—single particle motion and collective behavior: a review.
\newblock {\em Reports on Progress in Physics}, 78(5):056601, apr 2015.

\bibitem{Feng2024RoboticFish}
Haodong Feng, Dehan Yuan, Jiale Miao, Jie You, Yue Wang, Yi~Zhu, and Dixia Fan.
\newblock Efficient navigation of a robotic fish swimming across vortical flow fields.
\newblock {\em Journal of Hydrodynamics}, 36:1118–1129.

\bibitem{Frisch_1995}
Uriel Frisch.
\newblock {\em Turbulence: The Legacy of A. N. Kolmogorov}.
\newblock Cambridge University Press, Cambridge, 1995.

\bibitem{gupta2024flockingaidpathplanning}
Akanksha Gupta, Jaya~Kumar Alageshan, Kiran~Venkata Kolluru, and Rahul Pandit.
\newblock Can flocking aid the path planning of microswimmers in turbulent flows?
\newblock {\em Physics of Fluids}, 37(4):045107, 04 2025.

\bibitem{PhysRevE.100.053101}
Akanksha Gupta, Rohith Jayaram, Anando~G. Chaterjee, Shubhadeep Sadhukhan, Ravi Samtaney, and Mahendra~K. Verma.
\newblock Energy and enstrophy spectra and fluxes for the inertial-dissipation range of two-dimensional turbulence.
\newblock {\em Physical Review E}, 100(5):053101, November 2019.

\bibitem{manish2011targeted}
Manish Gupta and Vimukta Sharma.
\newblock Targeted drug delivery system: A review.
\newblock {\em Research Journal of Chemical Sciences}, 1(2):135--138, 2011.

\bibitem{Gustavsson_2017}
K.~Gustavsson, Luca Biferale, Antonio Celani, and Simona Colabrese.
\newblock Finding efficient swimming strategies in a three-dimensional chaotic flow by reinforcement learning.
\newblock {\em The European Physical Journal E}, 40(110), 2017.

\bibitem{Hussaini_1987}
M~Y Hussaini and T~A Zang.
\newblock Spectral methods in fluid dynamics.
\newblock {\em Annual Review of Fluid Mechanics}, 19(Volume 19, 1987):339--367, 1987.

\bibitem{Schneider_2011}
Benjamin Kadoch, Dmitry Kolomenskiy, Philippe Angot, and Kai Schneider.
\newblock A volume penalization method for incompressible flows and scalar advection–diffusion with moving obstacles.
\newblock {\em Journal of Computational Physics}, 231(12):4365--4383, 2012.

\bibitem{Kevlahan_2001}
Nicholas K.-R Kevlahan and Jean-Michel Ghidaglia.
\newblock Computation of turbulent flow past an array of cylinders using a spectral method with brinkman penalization.
\newblock {\em European Journal of Mechanics - B/Fluids}, 20(3):333--350, 2001.

\bibitem{kiverin2025twophasehydrodynamicmodelactive}
A.~Kiverin, S.~Luguev, and I.~Yakovenko.
\newblock Two-phase hydrodynamic model of active colloid motion, 2025.

\bibitem{koh2025physics}
Christopher Koh, Laurent Pagnier, and Michael Chertkov.
\newblock Physics-guided actor-critic reinforcement learning for swimming in turbulence.
\newblock {\em Physical Review Research}, 7(1):013121, 2025.

\bibitem{Kolomenskiy_2009}
Dmitry Kolomenskiy and Kai Schneider.
\newblock A fourier spectral method for the navier–stokes equations with volume penalization for moving solid obstacles.
\newblock {\em Journal of Computational Physics}, 228(16):5687--5709, 2009.

\bibitem{Kong2017RevisitingTM}
Xiangyu Kong, Bo~Xin, Fangchen Liu, and Yizhou Wang.
\newblock Revisiting the master-slave architecture in multi-agent deep reinforcement learning.
\newblock {\em arXiv preprint arXiv:1712.07305}, 2017.

\bibitem{Lauga2006}
Eric Lauga, W.~R. DiLuzio, George~M. Whitesides, and Howard~A. Stone.
\newblock Swimming in circles: Motion of bacteria near solid boundaries.
\newblock {\em Biophysical Journal}, 90(2):400--412, jan 2006.

\bibitem{Lesieur_2008}
Marcel Lesieur.
\newblock {\em Turbulence in Fluids}.
\newblock Springer, Dordrecht, 4 edition, 2008.

\bibitem{Li2009}
Guanglai Li and Jay~X. Tang.
\newblock Accumulation of microswimmers near a surface mediated by collision and rotational brownian motion.
\newblock {\em Phys. Rev. Lett.}, 103:078101, Aug 2009.

\bibitem{manzari2021targeted}
Mandana~T. Manzari, Yosi Shamay, Hiroto Kiguchi, Neal Rosen, Maurizio Scaltriti, and Daniel~A. Heller.
\newblock Targeted drug delivery strategies for precision medicines.
\newblock {\em Nature Reviews Materials}, 6(4):351--370, 2021.

\bibitem{mecanna2025critical}
Selim Mecanna, Aurore Loisy, and Christophe Eloy.
\newblock A critical assessment of reinforcement learning methods for microswimmer navigation in complex flows.
\newblock {\em The European Physical Journal E}, 48(10):58, 2025.

\bibitem{Michalec_2015}
François-Gaël Michalec, Sami Souissi, and Markus Holzner.
\newblock Turbulence triggers vigorous swimming but hinders motion strategy in planktonic copepods.
\newblock {\em Journal of The Royal Society Interface}, 12(106):20150158, 05 2015.

\bibitem{Mo2023IntelligentMicroswimmers}
Chaojie Mo, Gaojin Li, and Xin Bian.
\newblock Challenges and attempts to make intelligent microswimmers.
\newblock {\em Frontiers in Physics}, 11:1279883, 2023.

\bibitem{monthiller2022surfing}
R\'emi Monthiller, Aurore Loisy, Mimi A.~R. Koehl, Benjamin Favier, and Christophe Eloy.
\newblock Surfing on turbulence: A strategy for planktonic navigation.
\newblock {\em Physical Review Letters}, 129(6):064502, Aug 2022.

\bibitem{Mukherjee_2019}
Rupak Mukherjee, Akanksha Gupta, and Rajaraman Ganesh.
\newblock Compressibility effects on quasistationary vortex and transient hole patterns through vortex merger.
\newblock {\em Physica Scripta}, 94(11):115005, aug 2019.

\bibitem{Padhan_2024}
N.~B. Padhan, D.~Vincenzi, and R.~Pandit.
\newblock Interface-induced turbulence in viscous binary fluid mixtures, July 2024.
\newblock arXiv preprint arXiv:2407.13393.

\bibitem{padhan2025cahn}
Nadia~Bihari Padhan and Rahul Pandit.
\newblock The cahn--hilliard--navier--stokes framework for multiphase fluid flows: Laminar, turbulent and active.
\newblock {\em Journal of Fluid Mechanics}, 1010:P1, 2025.

\bibitem{Pandit_2017}
Rahul Pandit, Debarghya Banerjee, Akshay Bhatnagar, Marc Brachet, Anupam Gupta, Dhrubaditya Mitra, Nairita Pal, Prasad Perlekar, Samriddhi~Sankar Ray, Vishwanath Shukla, and Dario Vincenzi.
\newblock An overview of the statistical properties of two-dimensional turbulence in fluids with particles, conducting fluids, fluids with polymer additives, binary-fluid mixtures, and superfluids.
\newblock 29(11):111112, 10 2017.

\bibitem{pedley1992hydrodynamic}
T.~J. Pedley and J.~B. Kessler.
\newblock Hydrodynamic phenomena in suspensions of swimming microorganisms.
\newblock {\em Annual Review of Fluid Mechanics}, 24(1):313--358, 1992.

\bibitem{Qiu2022Symmetry}
Jingran Qiu, Navid Mousavi, Kristian Gustavsson, Chunxiao Xu, Bernhard Mehlig, and Lihao Zhao.
\newblock Navigation of micro-swimmers in steady flow: The importance of symmetries.
\newblock {\em Journal of Fluid Mechanics}, 932:A10, 2022.

\bibitem{Ramprasad_PRE2025}
Margam Ramprasad, Shubhadeep Mandal, and Pallab Sinha~Mahapatra.
\newblock Motion of a microswimmer in a lattice of obstacles: Effect of thermal fluctuations.
\newblock {\em Physical Review E}, 111(6):065402, Jun 2025.

\bibitem{Schneider_Stark_2019}
E.~Schneider and H.~Stark.
\newblock Optimal steering of a smart active particle.
\newblock {\em EPL (Europhysics Letters)}, 127(6):64003, November 2019.

\bibitem{schneider2005numerical}
Kai Schneider and Marie Farge.
\newblock Numerical simulation of the transient flow behaviour in tube bundles using a volume penalization method.
\newblock {\em Journal of Fluids and Structures}, 20(4):555--566, 2005.

\bibitem{singh2009nanoparticle}
Rajesh Singh and James~W. Lillard~Jr.
\newblock Nanoparticle-based targeted drug delivery.
\newblock {\em Experimental and Molecular Pathology}, 86(3):215--223, 2009.

\bibitem{Spagnolie2012}
Saverio~E. Spagnolie and Eric Lauga.
\newblock Hydrodynamics of self-propulsion near a boundary: Predictions and accuracy of far-field approximations.
\newblock {\em Journal of Fluid Mechanics}, 700:105--147, 2012.

\bibitem{Spagnolie2015}
Saverio~E. Spagnolie, Guillermo~R. Moreno-Flores, Denis Bartolo, and Eric Lauga.
\newblock Geometric capture and escape of a microswimmer colliding with an obstacle.
\newblock {\em Soft Matter}, 11(17):3396--3411, 2015.

\bibitem{10.1093/humupd/dmi047}
S.~S. Suarez and A.~A. Pacey.
\newblock Sperm transport in the female reproductive tract.
\newblock {\em Human Reproduction Update}, 12(1):23--37, 11 2005.

\bibitem{Sutton_1988}
R.~S. Sutton and A.~G. Barto.
\newblock Reinforcement learning: An introduction.
\newblock {\em IEEE Transactions on Neural Networks}, 9(5):1054--1054, 1998.

\bibitem{Tadmor_1991}
Eitan Tadmor.
\newblock Spectral methods in fluid dynamics (book review).
\newblock {\em Mathematics of Computation}, 57(196):876--878, 1991.

\bibitem{Tailleur_2008}
J.~Tailleur and M.~Cates.
\newblock Statistical mechanics of interacting run-and-tumble bacteria.
\newblock {\em Physical Review Letters}, 100(21):218103, 2008.

\bibitem{Takagi2013HydrodynamicCO}
Daisuke Takagi, J{\'e}r{\'e}mie Palacci, Adam~B. Braunschweig, Michael~J. Shelley, and Jun Zhang.
\newblock Hydrodynamic capture of microswimmers into sphere-bound orbits.
\newblock {\em Soft Matter}, 10(11):1784--1789, 2013.

\bibitem{Tan2024}
Rong Tan, Xiong Yang, Haojian Lu, and Yajing Shen.
\newblock One-step formation of polymorphous sperm-like microswimmers by vortex turbulence-assisted microfluidics.
\newblock {\em Nature Communications}, 15:4761, 2024.

\bibitem{verma_2020}
Mahendra~K. Verma, Abhishek Kumar, and Akanksha Gupta.
\newblock Hydrodynamic turbulence: Sweeping effect and taylor's hypothesis via correlation function.
\newblock {\em Transactions of the Indian National Academy of Engineering}, 5:649--662, 2020.

\bibitem{Volpe_2011}
Giovanni Volpe, Ivo Buttinoni, Dominik Vogt, Hans-Jürgen Kümmerer, and Clemens Bechinger.
\newblock Microswimmers in patterned environments.
\newblock {\em Soft Matter}, 7:8810--8815, 2011.

\bibitem{Xu_2020}
Tiantian Xu, Yanming Guan, Jia Liu, and Xinyu Wu.
\newblock Image-based visual servoing of helical microswimmers for planar path following.
\newblock {\em IEEE Transactions on Automation Science and Engineering}, 17(1):325--333, 2020.

\bibitem{Xue2024NanoMotors}
Jueyi Xue, Hamid Alinejad-Rokny, and Kang Liang.
\newblock Navigating micro- and nano-motors/swimmers with machine learning: Challenges and future directions.
\newblock {\em ChemPhysMater}, 3:273--283, 2024.

\bibitem{ZOU2024107666}
Zonghao Zou, Yuexin Liu, Alan C.~H. Tsang, Y.-N. Young, and On~Shun Pak.
\newblock Adaptive micro-locomotion in a dynamically changing environment via context detection.
\newblock {\em Communications in Nonlinear Science and Numerical Simulation}, 128:107666, 2024.

\end{thebibliography}
\end{document}